# Computing Longitudinal Dynamic Derivatives of a VTOL Aircraft Using CFD Simulations and Forced-Oscillation Model

Ali Khosravani Nezhad[1], AmirReza Kosari[2*], Rasoul Askari[3*]

1-MSc, University of Tehran, Collage of Interdisciplinary Science and Technologies, School of Aerospace Engineering, Ali.Khosravani@ut.ac.ir

2*-Associated Professor, University of Tehran, College of Interdisciplinary Science and Technologies, School of Aerospace Engineering. Kosari_a@ut.ac.ir

3-Assistant Professor, University of Tehran, College of Interdisciplinary Science and Technologies, School of Aerospace Engineering. Rasoul.askari@ut.ac.ir

**Abstract**

This study presents a comprehensive evaluation of dynamic aerodynamic derivatives during aircraft transition phases using advanced CFD simulations and forced oscillation testing. Two case studies are examined: a three-dimensional fighter aircraft (Standard Dynamic Model, SDM) and a UT24 eVTOL model. The transition phase—from vertical hover to forward cruise—is analyzed with harmonic oscillation techniques to capture unsteady aerodynamic forces and moments. Grid sensitivity studies and multi-zone meshing strategies ensure simulation accuracy, while ANSYS Fluent's finite volume solver and coupled pressure-velocity algorithms provide high-fidelity results. Dynamic derivatives are derived from variations in angle of attack, flight path, and rotational movements, with experimental and numerical data validating the approach. The findings offer valuable insights for robust control design and stability analysis, supporting future advancements in urban air mobility and aerospace engineering. Overall, this approach demonstrates substantial promise for optimizing aircraft performance during critical transition phases. These results pave the way for future innovations.

## 1. Introduction

The advancement of Vertical Take-Off and Landing (VTOL) aircraft has accelerated due to technological innovations and the increasing demand for flexible, efficient air transportation[1]. These aircraft can operate in diverse environments, including space-limited urban areas, but transitioning between hovering and forward flight presents significant challenges[2]. The transition phase is critical and complex, involving a shift from vertical hover to forward cruising, which profoundly affects aerodynamic performance, stability, and control [3]. During this phase, lift and drag forces dynamically change, and control authority shifts between the propulsion system and aerodynamic surfaces[4].

Traditional stability and control analysis methods, such as empirical models, exhibit significant limitations when applied to VTOL aircraft, particularly during the transition phase [5]. Originally developed for conventional fixed-wing aircraft, these models rely heavily on empirical data and semi-empirical formulas derived from wind tunnel tests of traditional airframes [6,7]. While effective for fixed-wing and rotary-wing aircraft under steady flight conditions, they are ill-suited to handle the complex, nonlinear aerodynamics of VTOL transitions [8]. Zaludin et al. [9] investigate the challenges and strategies for transitioning hybrid fixed-wing VTOL systems from hover to forward flight, addressing the shortcomings of empirical models. This paper proposes a hybrid fixed-wing VTOL and examines the associated transition challenges and strategies.

Empirical models inadequately capture the unique aerodynamic interactions during VTOL aircraft transitions from vertical hover to forward flight [5]. Tran et al. [10] utilized high-fidelity computational fluid dynamics (CFD) to examine wing-rotor interactions in the XV-15 tiltrotor during this transition, finding that rotor downwash increases wing lift by up to 13% at moderate to low nacelle angles but decreases lift and drag by 15% at high angles. Additionally, Chu et al. [11] developed a simulator in 2010 for VTOL micro air vehicle transition dynamics, creating a nonlinear longitudinal model based on wind tunnel data to better represent aerodynamic nonlinearities during flight mode changes.

The transition phase involves rapidly changing aerodynamic forces, complex flow separations, and interactions among multiple propulsion systems (e.g., distributed electric propulsion in eVTOLs) [12], which exceed DATCOM's empirical capabilities. DATCOM's steady-state assumptions fail to capture the unsteady aerodynamics of rapid transitions and lack the precision needed for modern electric and hybrid VTOL designs with novel airframes and distributed propulsion. Consequently, advanced methods like Computational

Fluid Dynamics (CFD) and experimental techniques such as the forced oscillation method are increasingly used to analyze aircraft dynamic derivatives [13]. These approaches provide enhanced accuracy by addressing nonlinearities, unsteady forces, and unique aerodynamic characteristics essential for contemporary VTOL designs.

Forced oscillation models provide a high-fidelity method for estimating an aircraft's dynamic derivatives, with numerous studies demonstrating their effectiveness. In 1983, Beyers [14] presented dynamic derivatives for angles of attack up to 40 degrees and sideslip angles between –5° and +5°, revealing significant nonlinear trends at high incidences and heightened sensitivity to sideslip angles. Subsequently, Balakrishna [15] (1987) conducted a dynamic wind tunnel study on the pitching moment coefficients of the Standard Dynamics Model, using flight test-like inputs and maximum likelihood estimation. The results showed strong agreement between flight test data and experimental outcomes.

This paper addresses the complexities of the VTOL transition phase by precisely identifying dynamic derivatives and introducing a novel integrated method that combines flight dynamics principles with Computational Fluid Dynamics (CFD). By merging CFD with the Forced Oscillation method, we establish a comprehensive framework that enhances the reliability of dynamic derivatives for innovative VTOL configurations. This integration facilitates a thorough understanding of dynamic behavior changes during transitions, ensuring accurate and robust dynamic modeling of VTOL aircraft throughout the design process.

## 2. Methodology

This study utilizes a Computational Fluid Dynamics (CFD) to simulate a reduced-order aircraft model with forced oscillation method coupled with multiple oscillations to mimic flight conditions. The resulting data allows for the decoupling of forces and moments, enabling the determination of dynamic derivatives and coefficients.

### 2.1. Forced Oscillation Method

Aircraft motion consists of six degrees of freedom, divided into longitudinal and lateral-directional forces and moments, characterized by four angles: Angle of Attack ($\alpha$), flight path angle ($q$). The forced oscillation method applies harmonic oscillations to induce dynamic behavior and monitor changes in forces and moments. By combining these oscillations, diverse movement patterns are created, isolating the pitch angle ($\theta$) from $\alpha$ and q through modulation, as shown in Figure 1.

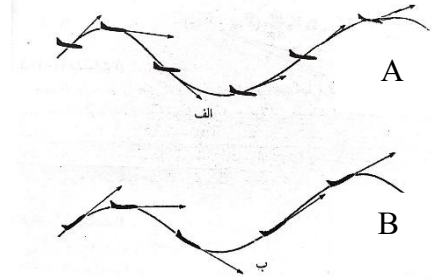

Figure 1. (A) Flight with q=0 and variable $\alpha$, (B) Flight with $\alpha = 0$ and variable q [15]

The harmonic oscillation equation is fundamental in dynamics, especially for systems undergoing simple harmonic motion (SHM). The general solution for simple harmonic motion is in the form:

$$x(t) = A\sin(\omega t + \phi) \qquad (1)$$

In the equation 1 the $A$ is the amplitude of the oscillation, $\omega$ is the angular frequency, and $\phi$ is the phase angle, which is determined by the initial conditions.

The longitudinal force on the airplane is the lift force, drag force, and pitching moment, this coefficient can be rewritten based on harmonic oscillation as below.

$$C_l(t) = C_{l_0} + C_{l_u}u(t) + C_{l_\alpha}A_\alpha \sin(\omega t + \phi_\alpha) \\ + C_{l_q}A_q\sin(\omega t + \phi_q) \qquad (2)$$

$$C_d(t) = C_{d_0} + C_{d_u}u(t) + C_{d_\alpha}A_\alpha \sin(\omega t + \phi_\alpha) \\ + C_{d_q}A_q\sin(\omega t + \phi_q) \qquad (3)$$

$$C_m(t) = C_{m_0} + C_{m_u}u(t) + C_{m_\alpha}A_\alpha \sin(\omega t + \phi_\alpha) + (C_{m_q} \\ + C_{m_{\dot\alpha}})A_q\cos(\omega t + \phi_q) \qquad (4)$$

In Equations 2–4, static coefficients (subscript 0) remain unaffected by transient motions $u$, $\alpha$, and $q$ represent changes due to forward speed, angle of attack, and flight path angle, respectively.

To evaluate dynamic derivatives separately, harmonic oscillations are applied to the aircraft at steady speed and constant airflow, eliminating the effects of forward speed and path angle. The dynamic coefficients for angle of attack are derived from the lift slope $(C_{l_\alpha})$, drag slope $(c_{d_\alpha})$, and pitching moment slope $(c_{m_\alpha})$ using equation 5 to 6.

$$C_l(t) = C_{l_0} + C_{l_\alpha}\alpha\sin(\omega t) \qquad (5)$$
$$C_d(t) = C_{d_0} + C_{d_\alpha}\alpha\sin(\omega t) \qquad (6)$$

$$C_m(t) = C_{m_0} + C_{m_\alpha}\alpha\sin(\omega t) \quad (7)$$

The same can be implemented to the flight path angle. It should be mentioned that the pitching moment coefficient related to flight path angle is coupled with the $c_{m_\alpha}$ which can be decoupled by rotating the body along the z-axis while constantly rotating it on the y-axis. This movement eliminates the change in the angle of attack the body faces relative to the free stream. Thus, the $c_{m_{\dot\alpha}}$ component is eliminated from the equation.

This modeling approach allows for a comprehensive examination of the combined oscillations for each mode, as elaborated in Table 1.

**Table 1. Harmonic oscillation is assigned to each axis to generate motion** [16]

| oscillation | Flight path mode $q$ | AoA mode $\alpha$ |
|---|---|---|
| Body x-axis | | |
| Body y-axis | $\alpha_B.\sin(\omega t)$ | $\alpha_B.\sin(\omega t)$ |
| Body z-axis | | |
| Flow x-axis | | |
| Flow y-axis | $\alpha_A.\sin(\omega t)$ | |
| Flow z-axis | | |

Based on the visual representation in Figure 1 and the equation provided, the intricate interplay of complex flow oscillation and rotational motion on the aircraft body is conceptualized through the application of harmonic oscillations in both the fluid flow direction and the aircraft body

### 2.2. Fluid Flow Numerical Model

The fluid flow physics in the presented study is dynamic and unsteady, with the Navier-Stokes equations functioning as the governing equations for compressible and unsteady flow. The current simulation is performed within a three-dimensional domain and discretized using a multi-zonal grid. Numerous investigations validate that the Unsteady Reynolds-Averaged Navier-Stokes (URANS) model satisfactorily captures tip vortices and turbulence interactions. For example, Miyaji et al. [17] Predicted the dynamic stability derivatives of aircraft via CFD methods, employing the RANS model in the JAXA solver for their simulation. In another study, Ghoreyshi et al., Ref [18], have examined the aerodynamic modeling of the SDM model to assess dynamic derivatives, utilizing forced oscillations in a reduced-order model and employing a RANS solver. The results demonstrate a strong correlation of data with a little error margin.

A sliding mesh is employed to simulate rotation to produce motion across the aircraft body, while the oscillations imposed on the bodies are executed using a UDF code. The discretization of the governing equations on the grid utilizes the cell-centered finite volume method, and to enhance the accuracy of gradient calculations, the least-squares method is used. The least-squares method reduces the discrepancy between anticipated and actual values of variables in adjacent cells. The pressure-velocity coupling system is configured as coupled, and all spatial discretization is established as second-order upwind. Based on the assumptions, the governing equations for fluid flow can be articulated as follows[19].

Unsteady compressible momentum equation:

$$\frac{\partial(\rho v)}{\partial t} + \nabla.(\rho v v) = -\nabla p + \nabla.(2\mu D) + \rho g \quad (8)$$

Mass conservation equation:

$$\frac{\partial \rho}{\partial t} + \nabla.(\rho v) = 0 \quad (9)$$

And the energy equation:

$$\frac{\partial \rho E}{\partial t} + \nabla.(\rho v E) = -\nabla.q + \rho W_s + \rho H_s \quad (10)$$

In the above equations, $\rho$ is the fluid density, $\vec{U}$ is the velocity vector, and $P$ Is the pressure.

This study used Generalized. $k - \omega$ (GEKO) Two-equation turbulence, which is recognized for its precision in capturing near-wall flow features [20]. The k-ω GEKO model, created by Menter in 2019 [21], integrates the advantageous characteristics of the $k$-$\varepsilon$ turbulence closure in free-stream areas with the reliability and precision of the model $k - \omega$ in near-wall regions. Incorporating transport factors into the eddy viscosity formulation enables the model to precisely forecast flow separation and adverse pressure gradients. The $k - \omega$ (GEKO) model employs two equations to determine the turbulent kinetic energy ($k$) and the length scale ($\omega$) for numerical problem-solving. Equations 11 and 12 denote the turbulent kinetic energy and the length scale, respectively[22].

$$\frac{\partial(\rho k)}{\partial t} + \frac{\partial(\rho u_i k)}{\partial x_i} = P_k - C_\mu \rho k \omega \\ + \frac{\partial}{\partial x_j}[(\mu + \frac{\mu_t}{\sigma_k})\frac{\partial k}{\partial x_j}] \quad (11)$$

$$\frac{\partial(\rho \omega)}{\partial t} + \frac{\partial(\rho u_i \omega)}{\partial x_i} = c_{\omega 1}F_1\frac{\omega}{k}P_k - c_{\omega 2}F_2\rho\omega^2 + \rho F_3 CD \\ + \frac{\partial}{\partial x_j}[(\mu + \frac{\mu_t}{\sigma_k})\frac{\partial \omega}{\partial x_j}] \quad (12)$$

In Equation 13, the left-hand side denotes the transient and convective components of turbulent kinetic energy ($k$ the right-hand side encompasses the diffusion, production, and dissipation components. Equation 14 has the initial three components on the right side and the two terms on the left side, which pertain to the disturbed frequency. This equation incorporates a cross-diffusion element, referred to as the source term, for the transition from $\varepsilon$ to $\omega$. Default model constants were utilized in ANSYS-Fluent 21.0 [20]. Also, with respect to the turbulence model, the thickness of the first layer for all tested grids was established at $1.56 \times 10^{-5}$ to achieve $y^+ = 1$ in wall boundaryes and capture the boundary layer.

## 3. Geometry & Mesh Independence

This study employs two case studies—the Standard Dynamic Model (SDM) and the UT24 eVTOL aircraft—to validate the FOM method and dynamic derivatives. The SDM, inspired by the F-16 fighter jet, serves as the reference geometry for grid resolution and simulation validation. It features a slender strake-delta wing, stabilizers, ventral fins, and a blocked inlet. Details are provided in Table 2 [23] and a three-view drawing in Figure 2 [23].

**Table 2 SDM geometry data**[23]

| Parameter | Value |
| --- | --- |
| $S\ (m^2)$ | 0.1238 |
| $b\ (m)$ | 0.6096 |
| $c\ (m)$ | 0.2299 |
| $d\ (m)$ | 0.9429 |

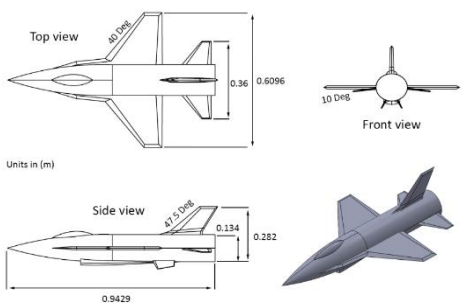

**Figure 2 SDM geometry used in this simulation and Dimensions of SDM model** [23]

The VTOL case study uses geometry adapted from the UT24 model, currently developed for eVTOL applications. Figure 3 illustrates the aircraft's dimensions and design, which features a Lift+Cruise configuration, V-tail, and a distributed electric propulsion system with eight rotors positioned along a wing-spanning boom. This configuration is optimized for Urban Air Mobility (UAM) missions.

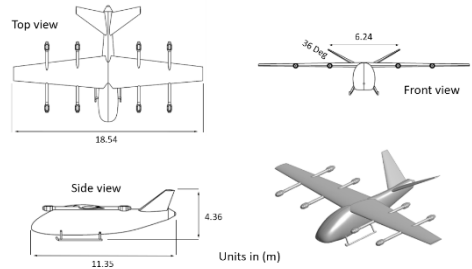

**Figure 3 UT24 Geometry and dimensions**

The presented geometries are used to perform CFD simulation with FOM applied to them. Which the details are discussed further.

### 2.3. Boundary conditions

This study simulates the SDM and eVTOL models by isolating the rotating domain (Figure 4). Domain B uses a cost-efficient structured grid and connects to sliding Domain A via an interface boundary. Multi-zone structured and unstructured grids accommodate rapid variations and diverse scenarios. The GEKO model maintains ($y^+ < +1$) through grid expansion, ensuring accurate boundary layer prediction with the selected turbulence model [20].

This study employs ANSYS Design Modeler for precise geometry creation and ANSYS Meshing for mesh generation. Fluid flow is simulated using finite volume-based ANSYS Fluent with a coupled pressure-velocity algorithm and second-order discretization for momentum, energy, and pressure equations. Unsteady simulations use a half-degree rotation time step and a first-order upwind scheme to accurately capture dynamics.

Aerodynamic coefficients are recorded to calculate dynamic derivatives, achieving convergence with residuals below $10^{-6}$ for continuity and momentum and below $10^{-8}$ for energy. Boundary conditions include a C-shaped domain for uniform inlet oscillations, domain extensions based on wingspan [18], and a rotating body extending 1.2 wingspans for robust data transfer. Pressure far-field conditions are applied to static domain sides, with harmonic oscillations managed via a UDF file. The interface enforces a no-slip condition with fixed rotation. All boundary conditions are illustrated in Figure 4.

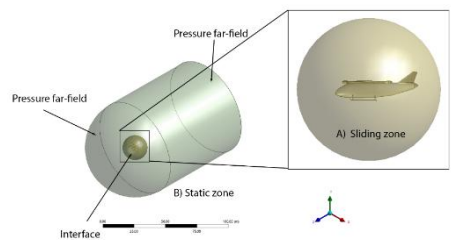

**Figure 4 Boundary condition assigned**

In the unsteady simulations, a solver time step corresponding to a one-degree rotation of the sliding domain was employed using a first-order upwind scheme to manage unsteady flows. Aerodynamic coefficients were continuously recorded, and perturbations were simplified with an averaged model for efficient analysis of complex interactions.

For the grid resolution study of the UT24 eVTOL aircraft, a multi-zone grid composed of tetrahedral and hexahedral elements with four different grid sizes was utilized. The primary parameters varied were the surface mesh element size for unstructured zones and the number of divisions for structured zones. A grid independence study revealed that a mesh with 6,219,531 cells reduced the difference in lift moment coefficients to approximately 5% compared to the finest mesh of 11,238,164 cells, demonstrating superior convergence and thus was selected for simulations.

Figure 5 illustrates the lift coefficient variations across different mesh sizes, showing decreased discrepancies with finer meshes. The final mesh grid applied to the aircraft and its surrounding domain is depicted in Figure 6.

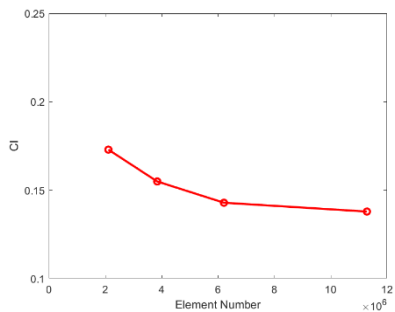

**Figure 5 Grid independency results over test grids**

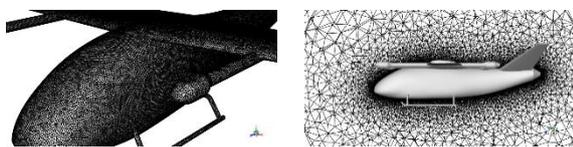

**Figure 6 left) mesh grid over case study body, right) mesh grid in domain**

## 4. Results & Discussion
### 4.1. Validation

Multiple test cases were used to validate the numerical method, with the primary case being a three-dimensional fighter aircraft (SDM). The RANS mesh is rectangular with the geometry centered, using no-slip adiabatic wall conditions at the body and a modified Riemann-invariant condition at the far field. The SDM was simulated at 100 m/s ($Re = 0.57 \times 10^6$) with α ranging from 0° to 70°, and the static force and moment coefficients were compared with experimental data [24] and Tatars' numerical data [23]. Figure 7 and 8 shows the resulting $C_n$ and $C_m$ values.

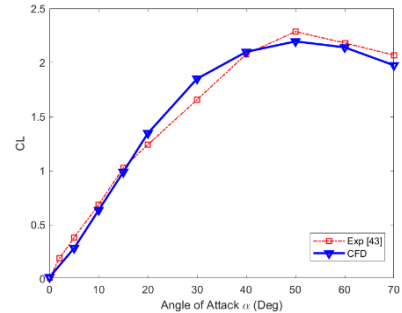

**Figure 7 SDM model validation**[23]

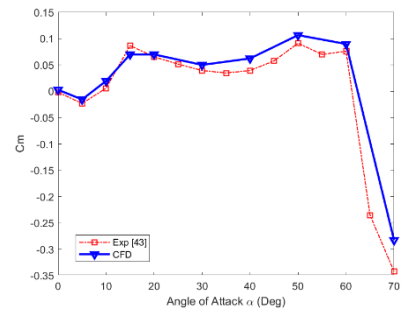

**Figure 8 SDM model validation**[23]

### 4.2. Results

The study aims to evaluate the dynamic derivatives during an aircraft's transition mode. Rigorous CFD simulations with forced oscillations were conducted to understand dynamic behavior, estimate derivative changes, support controller design, and enable future stability analyses. For a Lift+Cruise aircraft, the transition from hover/quadrotor to cruise, achieved through a combination of pusher and lift engine thrust, is divided into three scenarios start, middle, and end with Table 3 detailing the altitude, vertical, and forward velocities for each.

**Table 3 transition scenarios detail**

| Scenario | Altitude from ground | Vertical velocity | Forward Velocity |
|---|---|---|---|
| Transition beginning | 15 | 0 | 0 |
| Mid transition | 200 | 2.5 (m/s) | 33 (m/s) |
| Transition ends | 450 | 0 (m/s) | 66 (m/s) |

Table 3 outlines the transition scenarios within the extended mission profile of the eVTOL, as shown in Figure 9.

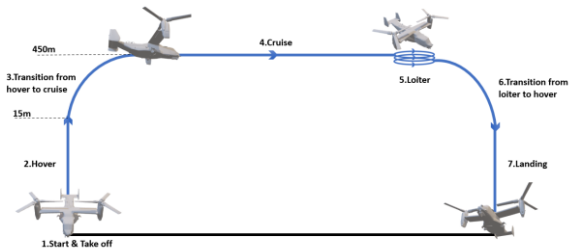

**Figure 9 eVTOL mission profile**

The reference reduced frequency oscillation is determined according to the values provided in Table 4.

**Table 4 Dynamic test condition AGARD CT2[25]**

| Parameter | Values |
| --- | --- |
| $M$ | 0.6 |
| Mean incidence $\alpha_0$ | 3.16 deg |
| Pitch Amplitude $\alpha_A$ | 4.59 deg |
| Reduced frequency $k$ | 0.0811 |

In the initial forced oscillation tests, the aircraft undergoes controlled angular displacements to capture dynamic derivatives related to angle-of-attack variations. This process reveals unsteady aerodynamic forces and moments across the transition phases (beginning, middle, and end). These derivatives dependent on AoA rate and angular velocities are key to understanding stability and control during complex transitions such as vertical to forward flight. Figure 10 to 13 uses arrows to indicate pitch-up and pitch-down motions, clarifying the relationship between oscillatory movement and aerodynamic response.

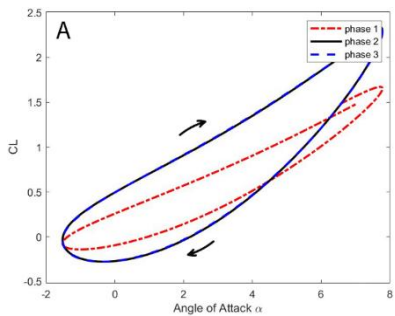

**Figure 10 Dynamic Lift coefficient with respect to angle of attack**

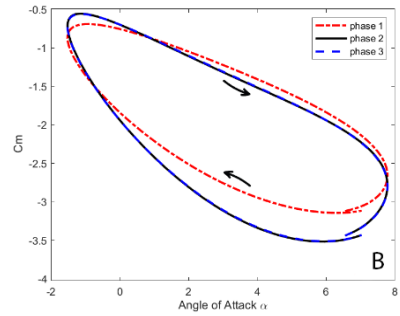

**Figure 11 Dynamic Pitch Moment with respect to change of angle of attack**

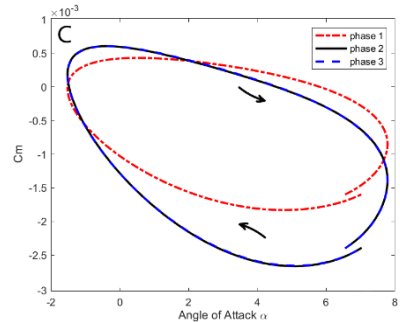

**Figure 12 Dynamic Pitch Moment with respect to change of angle of attack rate**

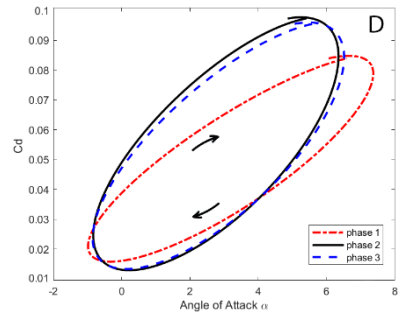

**Figure 13 Dynamic drag coefficient with respect to angle of attack**

The diagrams illustrate key aerodynamic changes during vertical flight transitions. Diagram in Figure 10 shows that the force coefficient increases with speed, with similar lift generation in mid and late transition. Diagram Figure 11 indicates a relatively constant rate of change for $C_{m_{\dot{\alpha}}}$ while Diagram Figure 12 shows minimal changes in the moment coefficient. Diagram Figure 13 reveals increased drag, evidenced by larger dynamic ellipses that indicate heightened flow separation during turn-down. Figure 14 and 15 correlates variations in lift $C_{L_q}$ and moment $C_{m_q}$ coefficients with changes in flight path angle, highlighting the influence of angle-of-attack shifts on stability and control.

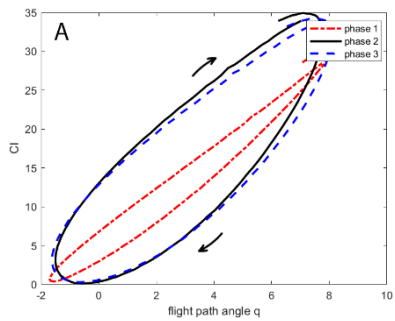

**Figure 14 Dynamic lift Coefficient with respect to change of flight path angle**

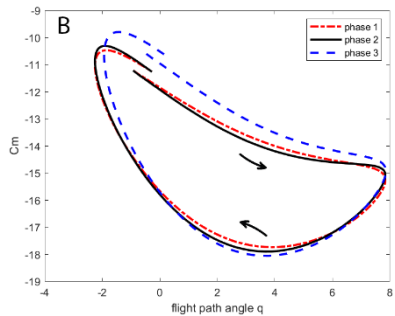

**Figure 15 Dynamic moment Coefficient with respect to change of flight path angle**

In Figure 14, the data illustrates a gradual but discernible increase in the lift coefficient as the forward speed of the aircraft increases. This increase is accompanied by a corresponding rise in the slope of the curve, indicating a proportional relationship between the lift coefficient and the forward speed. Shifting focus to graph Figure 15, it becomes apparent that the moments experienced in the middle and end of the transition phase are relatively similar. However, as the transition progresses, there is a noticeable increase in the curve slope, particularly towards the end of the transition. Furthermore, it is observed that a certain variable, denoted as $C_{m_q}$, Increases, there is a concurrent enhancement in the longitudinal stability of the aircraft.

## 5. Conclusion

In conclusion, this study successfully demonstrates a robust approach to evaluating dynamic aerodynamic derivatives during the transition phase of both conventional and eVTOL aircraft. By integrating advanced CFD simulations with forced oscillation testing, we captured the unsteady forces and moments that occur as aircraft transition from vertical to forward flight. The use of reduced-order models, exemplified by the SDM and UT24 eVTOL configurations, allowed for detailed analyses that are validated against experimental and numerical data. Grid sensitivity studies and precise meshing techniques further ensured the accuracy of our simulations, while the implementation of coupled pressure-velocity algorithms and high-order discretization schemes provided reliable insights into aerodynamic behavior. The results, illustrated through variations in lift, moment, and drag coefficients, confirm that the forced oscillation method is an effective tool for identifying dynamic derivatives crucial to stability and control. These findings not only contribute to the understanding of complex transitional aerodynamics but also lay the groundwork for the design of robust control strategies in future aircraft development, particularly for emerging urban air mobility applications.